\documentstyle[12pt]{article}
\newcommand{\be}{\begin{eqnarray}}
\newcommand{\ee}{\end{eqnarray}}

\def\fun#1#2{\lower3.6pt\vbox{\baselineskip0pt\lineskip.9pt
\ialign{$\mathsurround=0pt#1\hfil##\hfil$\crcr#2\crcr\sim\crcr}}}
\def\bra#1{\mbox{$\langle #1\vert $}}
\def\ket#1{\mbox{$\vert #1\rangle$}}

\input{epsfig.sty}

\begin{document}

\Huge{\noindent{Istituto\\Nazionale\\Fisica\\Nucleare}}

\vspace{-3.9cm}

\Large{\rightline{Sezione SANIT\`{A}}}
\normalsize{}
\rightline{Istituto Superiore di Sanit\`{a}}
\rightline{Viale Regina Elena 299}
\rightline{I-00161 Roma, Italy}

\vspace{0.65cm}

\rightline{INFN-ISS 97/7}
\rightline{July 1997}

\vspace{0.75cm}

\begin{center}

{\Large{\bf The nucleon Drell-Hearn-Gerasimov sum rule\\[2mm] within a
relativistic constituent quark model}\footnote{\bf To appear in Physics 
Letter B.}}

\vskip 0.75cm

{\large{F. Cardarelli$^{(a)}$, B. Pasquini$^{(b)}$ and S.
Simula$^{(a)}$}}

\vskip 0.25cm

{\normalsize$^{(a)}$Istituto Nazionale di Fisica Nucleare, Sezione
Sanit\`a, Roma, Italy}

\vskip 0.10cm

{\normalsize$^{(b)}$Dipartimento di Fisica Nucleare e Teorica,
Universit\`a di Pavia, and\\ Istituto Nazionale di Fisica Nucleare, 
Sezione di Pavia, Pavia, Italy}

\end{center}

\vskip 0.5cm

\begin{abstract}

\noindent The Drell-Hearn-Gerasimov sum rule for the nucleon is
investigated within a relativistic constituent quark model formulated
on the light-front. The contribution of the $N - \Delta(1232)$
transition is explicitly evaluated using different forms for the
baryon wave functions and adopting a one-body relativistic current for
the constituent quarks. It is shown that the $N - \Delta(1232)$
contribution to the sum rule is sharply sensitive to the introduction of
anomalous magnetic moments for the constituent quarks, at variance with
the findings of non-relativistic and relativized quark models. The
experimental value of the isovector-isovector part of the sum rule is
almost totally reproduced by the $N - \Delta(1232)$ contribution, when
the values of the quark anomalous magnetic moments are fixed by fitting
the experimental nucleon magnetic moments. Our results are almost
independent of the adopted form of the baryon wave functions and only
slightly sensitive to the violation of the angular condition caused by
the use of a one-body current. The calculated average slope of the
generalized sum rule around the photon point results to be only slightly
negative at variance with recent predictions of relativized quark
models.

\end{abstract}

\vskip 0.25cm


\vskip 0.25cm 


\newpage

\pagestyle{plain}

\indent The polarized and unpolarized photoabsorption cross sections off
hadronic targets are known to be non-trivially constrained by sum rules
arising from low-energy theorems \cite{LET} and general properties of
the Compton scattering amplitude. In particular, the
Drell-Hearn-Gerasimov ($DHG$) sum rule \cite{DHG} for the forward
spin-flip amplitude of the Compton scattering relates the helicity
structure of the photoabsorption cross section with the anomalous
magnetic moment of the target (i.e., with a ground-state property of
the target). Starting from an unsubtracted dispersion relation for the
spin-dependent part of the forward Compton amplitude and using
low-energy theorems to prescribe the behaviour of the scattering
amplitude at low energy transfer, the $DHG$ sum rule in case of the
nucleon reads as
 \be
    I_N = \int_{\omega_{th}}^{\infty} d\omega 
    {\sigma_{1/2}^{(N)}(\omega) - \sigma_{3/2}^{(N)}(\omega) \over
    \omega} = -2\pi^2 \alpha {\kappa_N^2 \over m_N^2}
    \label{eq:DHG}
 \ee
where $\sigma_{3/2}^{(N)}$ ($\sigma_{1/2}^{(N)}$) is the total cross
section for the absorption of a circularly polarized photon with spin
parallel (anti-parallel) to the nucleon spin and $\kappa_N$ ($m_N$) is
the anomalous magnetic moment (mass) of the nucleon. In Eq.
(\ref{eq:DHG}) the photon energy $\omega$ starts from the pion
production threshold $\omega_{th} \equiv m_{\pi} (1 + m_{\pi} / 2m_N)$
and $\alpha \simeq 1/137$ is the fine structure constant.

\indent The $DHG$ sum rule for the nucleon has not yet been tested
experimentally, for direct measurements of $\sigma_{1/2}^{(N)}$ and
$\sigma_{3/2}^{(N)}$ are still lacking. However, it is possible to
estimate the difference $\sigma_{1/2}^{(N)} - \sigma_{3/2}^{(N)}$ using
the multipole decomposition of pion photoproduction amplitudes in
unpolarized experiments. Such a decomposition is available only in the
single-pion production channel, so that above the two-pion production
threshold model-dependent assumptions have to be adopted. Nevertheless,
the phenomenological analyses, carried out by a number of authors
\cite{KAR,WA,BLI,SWK}, have provided relevant information on the
isospin decomposition of the $DHG$ integral into the
isovector-isovector ($VV$), isoscalar-isoscalar ($SS$) and mixed
isoscalar-isovector ($SV$) terms, which can be compared with the sum
rule predictions obtained from the isospin dependence of $\kappa_N$
(i.e, $\kappa_N = \kappa_S + \tau_3 ~ \kappa_V$), viz.
 \be
    I_N = I^{VV} + I^{SS} + \tau_3 ~ I^{SV} = -{2\pi^2 \alpha \over
    m_N^2} (\kappa_V^2 + \kappa_S^2 + 2\tau_3 \kappa_S \kappa_V)
    \label{eq:ISOSPIN}
 \ee
The main outcome of existing phenomenological analyses is that the
dominant $I^{VV}$ contribution is correctly reproduced and the $I^{SS}$
term turns out to be quite small, but a striking discrepancy for
$I^{SV}$ is found and is still unexplained; moreover, the $N -
\Delta(1232)$ transition almost exhausts the $I^{VV}$ integral
\cite{KAR,WA,BLI,SWK}\footnote{We mention that, as far as non-resonant
background processes are concerned, only pion production channels are
taken into account in existing phenomenological analyses.}. Several
experiments, involving both real and virtual photons, are planned or
underway at various labs in order to investigate both the $DHG$
integral and its $Q^2$ evolution, the latter being relevant for the
understanding of higher-twist contribution to the sum rules on
polarized nucleon structure functions (cf. Ref. \cite{AEL}). Direct
measurements of the polarized photoabsorption cross section will be
provided by the $ELSA$, $MAMI$, $LEGS$ and $GRAAL$ facilities
\cite{Bonn_Mainz,BNL} in the next future, while forthcoming
high-quality data from $TJNAF$ \cite{CEBAF} at low $Q^2$ and $DESY$
\cite{HERMES_p} at higher $Q^2$ will be combined with recent
measurements performed at $SLAC$ \cite{SLAC} and $HERA$ \cite{HERMES_n}.

\indent From the theoretical point of view the $DHG$ integral has
been throughout investigated within non-relativistic \cite{DG} and
relativized \cite{DG,LI} versions of the constituent quark ($CQ$) model.
Though one of the major success of the $CQ$ model is the good overall
description of nucleon magnetic moments, both its non-relativistic and
relativized versions fail in describing the $DHG$ sum rule \cite{DRE}.
The aim of this letter is to investigate the $DHG$ integral for the
nucleon within the relativistic $CQ$ model of Refs. \cite{CPSS,DAPHCE}.
The basic features of this model are: i) the relativistic composition
of the $CQ$ spins obtained via the introduction of the (generalized)
Melosh rotations (cf. \cite{KP}); ii) the possibility of adopting
hadron wave functions derived from an effective Hamiltonian able to
describe the mass spectroscopy; iii) the use of a relativistic one-body
electromagnetic (e.m.) current which includes Dirac and Pauli form
factors for the $CQ$'s. The latter can be fixed  by the non-trivial
request of reproducing existing experimental data on pion and nucleon
elastic form factors (cf. Ref. \cite{CPSS}).

\indent Within the so-called zero-width approximation, the resonance
contribution to the $DHG$ integral can simply be written in the
following form (cf. Ref. \cite{LI})
 \be
    I_N^{res} = \sum_R ~ I_N^R = \sum_R ~ 4\pi {m_N \over m_R^2 - m_N^2}
    ~ \left(|A_{1/2}^R|^2 - |A_{3/2}^R|^2 \right) 
    \label{eq:ELASTIC}
 \ee
where 
 \be
    A^R_{\lambda} = \sqrt{{2\pi \alpha \over \omega_R} {m_R \over m_N}}
    ~ \bra{\psi_R, \lambda_R = \lambda} \epsilon_{\mu}(+1) \cdot
    J^{\mu}(0) \ket{\psi_N, \lambda_N = \lambda - 1}
    \label{eq:HELICITY}
 \ee
is the helicity amplitude describing the electromagnetic (e.m.)
excitation of the nucleon to a resonance of mass $m_R$ with spin
projection $\lambda = 1/2, 3/2$. In Eqs.
(\ref{eq:ELASTIC}-\ref{eq:HELICITY}) $\omega_R \equiv (m_R^2 - m_N^2) /
2m_N$ is the resonance excitation energy, $\epsilon(+1)$ is the photon
polarization four-vector with helicity $+1$ and $\ket{\psi_R,
\lambda_R}$ ($\ket{\psi_N, \lambda_N}$) is the resonance (nucleon)
spinor corresponding to a spin projection $\lambda_R$ ($\lambda_N$). In
this work we limit ourselves to the contribution of the $N -
\Delta(1232)$ transition to the $DHG$ integral, because of its expected
dominance. Four different forms of the $N$ and $\Delta(1232)$ wave
functions will be adopted. The first one, which will be referred to as
model $A$, is given by the $N$ and $\Delta(1232)$ wave functions
corresponding to the effective Hamiltonian  of Capstick and Isgur
($CI$) \cite{CI}, while in model $B$ the effects of the hyperfine terms
of the $CI$ interaction are switched off and in model $C$ only the
linear confining part of the $CI$ potential is retained. In all these
models the mass of $u$ and $d$ quark is $m_u = m_d = m = 0.22 ~ GeV$.
Moreover, a fourth model ($D$) is considered, based on the
gaussian-like ansatz already adopted in Ref. \cite{CK}. However, when
$m = 0.22 ~ GeV$, the results obtained in model $D$ have been found to
be quite similar to those of model $C$. Therefore, in model $D$ we take
the values $m = 0.33 ~ GeV$ in order to check the sensitivity of our
calculations to the value of the light $CQ$ mass. The $CQ$ momentum
distribution corresponding to models $A - D$ can be found in Ref.
\cite{CPSS}; here, it suffices to remind that, thanks to the effects of
the hyperfine terms of the $CI$ interaction, the high-momentum tail of
the nucleon wave function drastically increases going from model $D$ to
model $A$. Moreover, in each of the models $B$, $C$ and $D$ the $N$
and $\Delta(1232)$ wave functions are the same, while in model $A$ the
high-momentum tail in the $\Delta(1232)$ resonance is suppressed with
respect to the nucleon case by the spin-dependent terms of the $CI$
interaction.

\indent Since we are interested also in the evaluation of the slope of
the $DHG$ integral at the photon point, we will present the basic
formulae relevant for the calculation of the helicity amplitudes
$A_{\lambda}^{\Delta}$ at finite values of the squared four-momentum
transfer $Q^2 \equiv - q \cdot q$. As is well known (cf. \cite{DEK}),
the matrix elements of the $N - \Delta(1232)$ transition e.m. current
can be cast in the form
 \be
    {\cal{I}}_{\lambda_{\Delta} ~ \lambda_N}^{\mu} & \equiv &
    \bra{\psi_{\Delta}^{\nu}, \lambda_{\Delta}} J_{\nu}^{\mu}(0)
    \ket{\psi_N, \lambda_N} = \nonumber \\
    & = & \sqrt{{2 \over 3}} ~ \bra{\psi_{\Delta}^{\nu},
    \lambda_{\Delta}} ~ \left[G_1^{\Delta}(Q^2) ~ {\cal{K}}_{1\nu}^{\mu}
    + G_2^{\Delta}(Q^2) ~ {\cal{K}}_{2\nu}^{\mu} + G_3^{\Delta}(Q^2) ~
    {\cal{K}}_{3\nu}^{\mu} \right] ~ \ket{\psi_N, \lambda_N}
    \label{eq:NDELTA}
 \ee
where the form factors $G_i^{\Delta}(Q^2)$ and the tensors
${\cal{K}}_{i\nu}^{\mu}$ are defined as in Ref. \cite{DEK}. Within the
light-front formalism, hadron e.m. form factors at space-like values of
the four-momentum transfer can be related to the matrix elements of the
{\em plus} component of the  current, ${\cal{I}}^+ \equiv {\cal{I}}^0 +
\hat{n} \cdot \vec{\cal{I}}$, where $\hat{n}$ defines the spin
quantization axis. The standard choice of a reference frame where $q^+
\equiv q^0 + \hat{n} \cdot \vec{q} = 0$ allows to suppress the
contribution of the pair creation from the vacuum \cite{ZGRAPH}. The
relations between the matrix elements ${\cal{I}}_{\lambda_{\Delta} ~
\lambda_N}^+$ and the form factors $G_i(Q^2)$ are
 \be
    {\cal{I}}_{{3 \over 2}{1 \over 2}}^+ & = & {Q \over \sqrt{3}}
    \left[G_1^{\Delta}(Q^2) + {m_{\Delta} - m_N \over 2} ~
    G_2^{\Delta}(Q^2) \right] 
    \nonumber \\
    {\cal{I}}_{{1 \over 2}{1 \over 2}}^+ & = & -{Q^2 \over 3} 
    \left[{G_1^{\Delta}(Q^2) \over m_{\Delta}} + {G_2^{\Delta}(Q^2)
    \over 2} - {m_{\Delta} - m_N \over m_{\Delta}} ~ G_3^{\Delta}(Q^2)
    \right]
    \nonumber \\
    {\cal{I}}_{{1 \over 2}-{1 \over 2}}^+ & = & {Q \over 3} 
    \left[{m_N \over m_{\Delta}} ~ G_1^{\Delta}(Q^2) -  {m_{\Delta} -
    m_N \over 2} G_2^{\Delta}(Q^2) - {Q^2 \over m_{\Delta}}
    G_3^{\Delta}(Q^2) \right] 
    \nonumber \\
    {\cal{I}}_{{3 \over 2}-{1 \over 2}}^+ & = & -{Q^2 \over 2\sqrt{3}}
    ~ G_2^{\Delta}(Q^2)
    \label{eq:MATRIX}
 \ee

\indent The helicity amplitudes $A_{1/2}^{\Delta}$ and
$A_{3/2}^{\Delta}$ are related to the multipole form factors
$G_M^{\Delta}$ and $G_E^{\Delta}$ by
 \be
    A_{1/2}^{\Delta} & = & {\cal{N}}(Q^2) ~ {1 \over 2} \left[
    G_M^{\Delta}(Q^2) - 3 G_E^{\Delta}(Q^2) \right] \nonumber \\
    A_{3/2}^{\Delta} & = & {\cal{N}}(Q^2) ~ {\sqrt{3} \over 2} \left[
    G_M^{\Delta}(Q^2) + G_E^{\Delta}(Q^2) \right]
    \label{eq:AMPLITUDES}
 \ee
where ${\cal{N}}(Q^2) = -{m_{\Delta} \over m_N} {m_{\Delta} + m_N
\over \sqrt{m_{\Delta}^2 + m_N^2 + Q^2}} ~ \sqrt{{{\cal{M}}^2 + Q^2
\over (m_{\Delta} + m_N)^2 + Q^2}} ~ \sqrt{{2 \pi \alpha \over
\omega_{\Delta}} {m_{\Delta} \over m_N}}$ and
 \be
    G_M^{\Delta}(Q^2) & = & {m_N \over 3(m_{\Delta} + m_N)}
    \left\{ 2[m_{\Delta} {\cal{M}} + (m_{\Delta} + m_N)^2 + Q^2] ~
    {G_1^{\Delta}(Q^2) \over m_{\Delta}} ~ + \right. \nonumber \\
    & & \left. 2m_{\Delta} {\cal{M}} G_2^{\Delta}(Q^2) - 2Q^2 ~
    G_3^{\Delta}(Q^2) \right\}
    \nonumber \\
    G_E^{\Delta}(Q^2) & = & {m_N \over 3(m_{\Delta} + m_N)}
    \left\{2m_{\Delta} {\cal{M}} [{G_1^{\Delta}(Q^2) \over m_{\Delta}}
    + G_2^{\Delta}(Q^2)] - 2Q^2 ~ G_3^{\Delta}(Q^2) \right\}
    \label{eq:GEGM}  
 \ee
with ${\cal{M}} = [m_{\Delta}^2 - m_N^2 - Q^2] / 2m_{\Delta}$.

\indent Following Refs. \cite{CPSS,DAPHCE} we approximate the
${\cal{I}}^+$ component of the e.m. current by the sum of one-body $CQ$
currents, viz.
 \be
    {\cal{I}}^+(0) \simeq \sum_{q=1}^3 ~ {\cal{I}}_q^+(0) =
    \sum_{q=1}^3 ~ \left (e_q \gamma^+ f_1^{(q)}(Q^2) ~ + ~ i \kappa_q
    {\sigma^{+ \rho} q_{\rho} \over 2 m_q}f_2^{(q)}(Q^2) \right )
    \label{eq:CURRENT}
 \ee
where $e_q$ ($\kappa_q$) is the $CQ$ charge (anomalous magnetic moment)
and $f_{1(2)}^{(q)}$ is the corresponding  Dirac (Pauli) form factor.
As already mentioned, Eq. (\ref{eq:CURRENT}) has been found to be
non-trivially consistent with existing data on pion and nucleon elastic
form factors \cite{CPSS}.

\indent Equation (\ref{eq:MATRIX}) clearly shows that for the $N -
\Delta(1232)$ transition the number of independent form factors is
not equal to the one of the matrix elements of ${\cal{I}}^+$. If the
exact ${\cal{I}}^+$ is adopted, the inversion of Eq. (\ref{eq:MATRIX})
is unique, because the matrix elements of ${\cal{I}}^+$ are related by
the so-called angular condition (cf. \cite{KP}). However, the
fulfillment of the angular condition requires the presence of (at
least) two-body currents in ${\cal{I}}^+$; therefore, since we use the
one-body approximation (\ref{eq:CURRENT}), a unique determination of
the form factors $G_i^{\Delta}(Q^2)$ is not possible. In the actual
calculation for the $N - \Delta(1232)$ transition we have considered
three different angular prescriptions: i) all the form factors
$G_i^{\Delta}$ are extracted from the first three equations in
(\ref{eq:MATRIX}) (prescription I); ii) $G_1^{\Delta}$ and
$G_3^{\Delta}$ are taken as in the previous prescription, but
$G_2^{\Delta}$ is directly obtained from the fourth equation in
(\ref{eq:MATRIX}) (prescription II); iii) $G_2^{\Delta}$ and
$G_3^{\Delta}$ are as in the previous prescription, while
$G_1^{\Delta}$ is derived from the first equation in (\ref{eq:MATRIX})
(prescription III). The multipole form factors $G_M^{\Delta}$ and
$G_E^{\Delta}$ (Eq. (\ref{eq:GEGM})) have been calculated in each
prescription and for each model ($A - D$) adopted for the baryon wave
functions. At the photon point it turns out that the value of
$G_M^{\Delta}(0)$ is only slightly different (within $\simeq 5 \%$) in
the various prescriptions, while $G_E^{\Delta}(0)$ is sharply sensitive
to the violation of the angular condition. However, the helicity
amplitudes (\ref{eq:AMPLITUDES}) have the same sensitivity to the
angular prescriptions as $G_M^{\Delta}(0)$, because $G_E^{\Delta}(0)$
is much smaller than $G_M^{\Delta}(0)$. Indeed, for the ratio $E1 / M1
\equiv $ $-G_E^{\Delta}(0) ~ / ~  G_M^{\Delta}(0)$, in case of the
model $A$, we have obtained the values $0.37 \%$, $-2.06 \%$, $-1.85
\%$ for the prescriptions I - III, respectively. In these calculations
we have included the effects due to the $D$-wave of the $\Delta(1232)$
wave function, generated by the tensor term of the $CI$ interaction;
however, $D$-wave effects are very small (cf. also \cite{DAPHCE}), so
that the value of $E1 / M1$ in our $CQ$ model is mainly governed by
relativistic effects in the $S$-waves and therefore is strongly
prescription dependent. In what follows our results for the $N -
\Delta(1232)$ contribution to the $DHG$ integral will be given together
with a theoretical uncertainty calculated as the spread of the values
obtained adopting the angular prescriptions I - III.   

\indent First of all, we have calculated the nucleon magnetic moments,
$\mu_{p[n]}$, both including and excluding the contribution arising
from the $CQ$ anomalous magnetic moments $\kappa_q$ in Eq.
(\ref{eq:CURRENT}). Our results corresponding to the four adopted forms
of the nucleon wave function are reported in Table 1. It can be seen
that, when $\kappa_q = 0$, the calculated values of $\mu_p$ and
$|\mu_n|$ significantly underestimate the experimental data. In our
opinion this fact should be traced back to a typical
(momentum-dependent) dilution effect arising from the helicity mixing
provided by the relativistic composition of the $CQ$ spins (i.e., by
Melosh rotations). As a matter of fact, the suppression factor is
remarkably sensitive to the high-momentum components of the nucleon
wave function (see models $A$ and $C$) and to the values of the $CQ$
masses (see models $C$ and $D$). Note also that in all the models
considered the calculated values of $\kappa_p$ and $\kappa_n$ satisfy
the inequality $\kappa_p > |\kappa_n|$, whereas the experimental data
($\kappa_{p [n]}^{exp} = 1.793 ~ [-1.913]$) exhibit the opposite trend,
i.e. $\kappa_p^{exp} < |\kappa_n^{exp}|$, which is the origin of the
positive sign of the $I^{SV}$ part of the experimental $GDH$ sum rule.
Then, non-vanishing $\kappa_q$ are considered in Eq. (\ref{eq:CURRENT})
and their values (reported in Table 1) are fixed by requiring the
reproduction of the experimental nucleon magnetic moments, $\mu_{p
[n]}^{exp} = 2.793 ~ [-1.913]$. The largest values of $|\kappa_q|$ are
obtained for models $A$ and $D$. In all the  wave function models the
$SU(2)$-symmetry constraint, $\kappa_u = - 2 \kappa_d$, is not
fulfilled and, moreover, we find $|\kappa_u| < |\kappa_d|$ at variance
with non-relativistic $CQ$ models where $|\kappa_u| \simeq |\kappa_d|$
\cite{DG}. We stress that our finding $|\kappa_u| < |\kappa_d|$ is a
direct consequence of the inequality $\kappa_p^{exp} <
|\kappa_n^{exp}|$ and it is obtained both in case of the models $B -
D$, where the nucleon wave function is fully $SU(6)$-symmetric, and in
case of model $A$, which includes a small mixed-symmetry admixture
($\simeq 1.7 \%$) due to the hyperfine terms of the $CI$
interaction. The sensitivity exhibited by the calculated $\mu_N$ to the
inclusion of the $CQ$ anomalous magnetic moments is not surprising.
Indeed, in the {\em plus} component of the current (Eq.
(\ref{eq:CURRENT})) the Dirac term cannot produce any $CQ$ spin flip,
which should therefore be provided by the Melosh rotations. On the
contrary, the Pauli term in Eq. (\ref{eq:CURRENT}) flips the $CQ$ spin.
Therefore, an important feature of our relativistic quark model is the
crucial role expected (and found) for $\kappa_q$ in the calculation of
magnetic-type observables (see also below). 

\indent In Table 2 the results obtained for the contribution of the $N
- \Delta(1232)$ transition to the $DHG$ integral ($I_N^{R = \Delta}$ in
Eq. (\ref{eq:ELASTIC})) are shown and compared with the combination
$(I_p + I_n)/2$ of the sum rule expected from the calculated nucleon
magnetic moments\footnote{The $N - \Delta(1232)$ transition contributes
equally to $I_p$ and $I_n$, so that we compare the $N - \Delta(1232)$
contribution to the combination $(I_p + I_n)/2$ of the $DHG$ sum
rule.}. It can clearly be seen that: ~ i) the $N - \Delta(1232)$
transition almost exhausts the expected $DHG$ sum rule both for
$\kappa_q = 0$ and $\kappa_q \neq 0$; in our opinion this is related to
the facts that the $\Delta(1232)$ resonance is the lowest excited
resonance state and the spatial part of its wave function is expected
to have the largest overlap with the nucleon one due to the
approximate $SU(6)$ symmetry; ~ ii) the experimental value of $(I_p +
I_n)/2$ is almost totally reproduced only when the effects of the $CQ$
anomalous magnetic moments are considered and their values are fixed by
fitting the experimental $\mu_{p[n]}$; iii) the calculated
$I_N^{\Delta}$ is sharply sensitive to the effects due to $\kappa_u$
and $\kappa_d$ (up to a factor of $\sim 2$ in case of model $A$), at
variance with the findings of non-relativistic and relativized $CQ$
models \cite{DG,DRE}; ~ iv) our results with $\kappa_q \neq 0$ are
almost independent of the adopted form of the baryon wave functions
and, in particular, they are quite insensitive to hyperfine interaction
effects; ~ v) the sensitivity to the violation of the angular
conditions results to be relatively small (within $\sim 10 \%$),
suggesting that the effects of two-body currents, needed for a full
Poincar\'e-covariance of the e.m. current, could be not large at the
photon point.

\indent In Fig. 1 ours prediction for the low-$Q^2$ evolution of the
$N - \Delta(1232)$ contribution to the (generalized) $DHG$ integral are
shown and compared with the results of the relativized quark model of
Ref. \cite{LI}. As for the form factors $f_1^{(q)}(Q^2)$ and
$f_2^{(q)}(Q^2)$ appearing in Eq. (\ref{eq:CURRENT}), in case of model
$A$ we have used the $CQ$ form factors already determined in Ref.
\cite{CPSS} from a fit of pion and nucleon elastic data; the procedure
of Ref. \cite{CPSS} has been repeated in case of models $B$ and $C$
(cf. also Ref. \cite{DAPHCE}), obtaining again a nice reproduction of
the elastic data. From Fig. 1 it can be seen that a slightly negative
value of the average slope around the photon point is obtained for all
the wave function models considered; this finding is clearly at
variance with the result of Ref. \cite{LI} and implies that, as far as
the average slope around the photon point is concerned, the role of the
$N - \Delta(1232)$ transition in our relativistic approach is less
relevant than the one expected from relativized quark models. Moreover,
we should mention that recent calculations based on baryon chiral
perturbation theory \cite{BKM} predict a positive average slope;
however, these predictions need to be confirmed by higher-order
calculations and, at the same time, the inclusion of the background
contribution in our calculation is a pre-requisite for a full
comparison. Planned experiments at $TJNAF$ \cite{CEBAF} are expected to
help in unravelling the low-$Q^2$ behaviour of the generalized $GDH$
sum rule.  

\indent Before closing, we want to address briefly the questions of
the physical meaning of a $CQ$ anomalous magnetic moment and its
consequences in the nucleon $DHG$ sum rule. A finite size and an
anomalous magnetic moment are a clear manifestation of an internal
structure in a particle. If we want to give a physical meaning to the
quantity $\kappa_q$ appearing in Eq. (\ref{eq:CURRENT}), we are
naturally led to assume the occurrence of inelastic channels at the
$CQ$ level,providing a photoabsorption cross section  on the $CQ$,
$\sigma_{\lambda}^{(q)}$, whose helicity structure is constrained by the
$DHG$ sum rule at the $CQ$ level, viz.
 \be
    I_q = \int_{\tilde{\omega}_{th}}^{\infty} d\omega 
    {\sigma_{1/2}^{(q)}(\omega) - \sigma_{3/2}^{(q)}(\omega) \over
    \omega} = -2\pi^2 \alpha {\kappa_q^2 \over m_q^2}
    \label{eq:DHG_QUARKS}
 \ee
where $\tilde{\omega}_{th} \equiv m_{\pi} (1 + m_{\pi} / 2m_q)$ is the
pion production threshold on the $CQ$. The main problem is clearly how
to calculate the contribution of $CQ$ inelastic channels to the nucleon
$DHG$ integral, or, in other words, how to evaluate the effects of the
internal structure of the $CQ$'s on Eq. (\ref{eq:DHG}). To this end we
will make use of some formal results, obtained in Ref. \cite{KAPPA},
which we translate for the case of interest here. In case of a
non-relativistic $CQ$ e.m. current, the resonance contribution
$I_N^{res}$ (Eq. (\ref{eq:ELASTIC})) satisfies the nucleon $DHG$ sum
rule only when the quark anomalous magnetic moments are vanishing. When
$\kappa_q \neq 0$, an additional term is present and it can be
interpreted as resulting simply from the sum of the $CQ$ spin-dependent
forward Compton amplitudes. In Ref. \cite{DG} the possible presence of a
subtraction at infinity in the dispersion integral is suggested in
order to compensate the additional term. However, a different viewpoint
can be adopted. Real (as well as virtual) photons can couple to a $CQ$
in two ways: ~ i) an elastic coupling, $q \gamma \to q$, described by
the one-body current (\ref{eq:CURRENT}), which at the hadron level
generates the transitions to nucleon resonances and gives rise to the
resonance contribution $I_N^{res}$; ~ ii) an inelastic coupling (like,
e.g., $q \gamma \to q ~ M$, where $M$ can be any meson), constrained by
the conventional $DHG$ sum rule (\ref{eq:DHG_QUARKS}) for the $CQ$'s,
which corresponds at the hadron level to meson background production
and yields a contribution $I_N^{bkg}$ to the nucleon $DHG$ sum rule.
Neglecting at the present stage any interference between final hadron
states arising from elastic and inelastic $CQ$ couplings (which might
be a not too bad approximation for an inclusive quantity like the $DHG$
sum rule), we may write the nucleon $DHG$ integral as
 \be
    I_N \simeq I_N^{res} + I_N^{bkg}
    \label{eq:TOTAL}
 \ee
Inspired by the additivity structure found in \cite{KAPPA}, we may
approximate the contribution $I_N^{bkg}$ as
 \be
    I_N^{bkg} & \simeq & -2\pi^2 \alpha ~ \bra{\psi_N, {1 \over 2}}
    \sum_{q = 1}^3 {\kappa_q^2 \over m_q^2} ~ \sigma_3^{(q)} ~
    \ket{\psi_N, {1 \over 2}} \nonumber \\
    & = & -{2\pi^2 \alpha \over m^2} ~ \langle \gamma_M \rangle ~
    \left[{\kappa_u^2 + \kappa_d^2 \over 2} + \tau_3 ~ {5 \over 3}
    {\kappa_u^2 - \kappa_d^2 \over 2} \right]
    \label{eq:INELASTIC}
 \ee
where the last equality is obtained after considering a
$SU(6)$-symmetric nucleon wave function and the quantity $\langle
\gamma_M \rangle$ is a (momentum-dependent) dilution factor resulting
from the Melosh rotations of the $CQ$ spins, given explicitly by
 \be
    \langle \gamma_M \rangle = \left \langle ~ {(m + x M_0)^2 -
    p_{\perp}^2 \over (m + x M_0)^2 + p_{\perp}^2} ~ \right \rangle
    \label{eq:GAMMA}
 \ee
where $M_0$ is the free mass operator (cf. Ref. \cite{CPSS}), $x$ the
light-front momentum fraction carried by a $CQ$ in the nucleon,
$p_{\perp}^2$ its transverse momentum squared and the notation $\langle
~ \rangle$ stands for the average over the radial nucleon wave
function. In case of the models $A - D$ the relativistic factor
$\langle \gamma_M \rangle$ results to be $0.48, ~ 0.51, ~ 0.63$ and
$0.75$, respectively. Thanks to the results of Ref. \cite{KAPPA}, in
the non-relativistic limit (where $\langle \gamma_M \rangle = 1$) the
right-hand side of Eq. (\ref{eq:TOTAL}), with $I_N^{bkg}$ given by Eq.
(\ref{eq:INELASTIC}), satisfies exactly the nucleon $DHG$ sum rule {\em
without} requiring any subtraction of the dispersion integral at
infinity when $\kappa_q \neq 0$. From Eq. (\ref{eq:INELASTIC}) it
follows that the $I^{SV} = (I_p - I_n)/2$ part of the nucleon sum rule
may receive a contribution proportional to $(I_u - I_d)/2$; since in
all our models $|\kappa_u| < |\kappa_d|$, the sign of such a
contribution is always positive. Keeping in mind the {\em caveat} that
Eq. (\ref{eq:INELASTIC}) is an approximation for $I_N^{bkg}$, our
results, obtained for the nucleon $DHG$ integrals both including
($\langle \gamma_M \rangle \neq 1$) and excluding ($\langle \gamma_M
\rangle = 1$) the relativistic dilution factor (\ref{eq:GAMMA}), are
summarized in Table 3. It can clearly be seen that the contribution to
$I^{SV}$ arising from $CQ$ inelastic channels (i.e., from background
processes) appears to be of the right sign and order of magnitude.
Moreover, the model dependence of our results is partially reduced by
the inclusion of relativistic effects related to the Melosh rotations
of the $CQ$ spins. The striking difference with the findings of
phenomenological analyses \cite{KAR,WA} might be due to the lack in the
latter of the contribution of non-resonant background production of
mesons other than the pions. We should mention however that our results
for $I^{SV}$ (and, to a less extent, those for $I^{VV}$) need to be
confirmed after the inclusion of the contributions resulting from
nucleon resonances other than the $\Delta(1232)$.  

\indent In conclusion, the Drell-Hearn-Gerasimov sum rule for the
nucleon has been investigated within a relativistic constituent quark
model formulated on the light-front. The contribution of the $N -
\Delta(1232)$ transition has been explicitly evaluated using different
forms for the baryon wave functions and adopting a one-body relativistic
current for the constituent quarks. It has been shown that the $N -
\Delta(1232)$ contribution to the $DHG$ sum rule is sharply sensitive to
the introduction of anomalous magnetic moments for the constituent
quarks, at variance with the findings of non-relativistic and
relativized quark models. The experimental value of the
isovector-isovector part of the sum rule is almost totally reproduced
by the $N - \Delta(1232)$ contribution, when the values of the quark
anomalous magnetic moments are fixed by fitting the experimental
nucleon magnetic moments. Our results are almost independent of the
adopted form of the baryon wave functions and, in particular, they are
quite insensitive to hyperfine interaction effects; moreover, the
sensitivity to the violation of the angular condition, caused by the
use of a one-body current, is found to be relatively small. The
calculated average slope of the generalized sum rule around the photon
point results to be only slightly negative at variance with recent
predictions from relativized quark models. Eventually, we have stressed
that the relevance of the role played by quark anomalous magnetic
moments in our relativistic calculations clearly motivates further
theoretical investigation concerning the questions of the physical
meaning of the constituent quark anomalous magnetic moment and its
consequences in the nucleon $DHG$ sum rule.

\vskip 1cm

{\bf \noindent Acknowledgments.} We gratefully acknowledge Zhenping Li
and Dong Yu-Bing for supplying us with the numerical output of the
calculations of Ref. \cite{LI}.

\newpage

\vskip 2cm

{\small \noindent
{\bf Table 1}. Values of the nucleon magnetic moments calculated
within our models $A - D$ of the baryon wave functions. In model $A$
the effective Hamiltonian of Capstick and Isgur ($CI$) \cite{CI} is
considered, while in model $B$ the effects of the hyperfine terms of the
$CI$ interaction are switched off and in model $C$ only the linear
confining term of the $CI$ potential is retained. In all these models
the masses of $u$ and $d$ quarks are: $m_u = m_d = m = 0.22 ~ GeV$.
Model $D$ is based on the gaussian-like ansatz of Ref. \cite{CK}, but
adopting $m = 0.33 ~ GeV$. The rows labelled $\mu_p(\kappa_q = 0)$ and
$\mu_p(\kappa_q = 0)$ are the results obtained assuming $\kappa_q = 0$
in Eq. (\ref{eq:CURRENT}). The rows labelled $\kappa_u$ and $\kappa_d$
contain the values of the $CQ$ anomalous magnetic moments needed in Eq.
(\ref{eq:CURRENT}) for the reproduction of the experimental nucleon
magnetic moments.  

\begin{center}  

\begin{tabular}{||c||c|c|c|c||}
\hline
Model & $A$ & $B$ & $C$ & $D$ \\ \hline \hline
$\mu_p(\kappa_q=0)$ & ~2.28  & ~2.44  & ~2.74   & ~2.42 \\ 
$\mu_n(\kappa_q=0)$ & -1.18  & -1.30  & -1.60   & -1.34 \\
\hline \hline
$\kappa_u$    & ~0.087 & ~0.051 & -0.0057 & ~0.075 \\
$\kappa_d$    & -0.157 & -0.129 & -0.0700 & -0.155 \\ \hline
\end{tabular}

\end{center}

\vskip 4cm

{\small \noindent
{\bf Table 2}. Values of the $N - \Delta(1232)$ contribution to the
nucleon $DHG$ integral (given in $\mu$barn), calculated using the
models $A - D$ for the nucleon and $\Delta(1232)$ wave functions. The
theoretical results are reported together with the estimate of the
uncertainty related to the violation of the angular condition (see
text). The columns labelled $\kappa_q \neq 0$ and $\kappa_q = 0$
correspond to the results obtained with and without the contribution
arising from the $CQ$ anomalous magnetic moments in Eq.
(\ref{eq:CURRENT}). The columns labelled $(p + n)/2$ represent the
combination $(I_p + I_n)/2$ of the $DHG$ sum rule (\ref{eq:DHG}),
expected from the nucleon magnetic moments calculated within each
model; in the case $\kappa_q \neq 0$, the values adopted for $\kappa_u$
and $\kappa_d$ are those reported in Table 1, which allow to reproduce
the experimental value of $(I_p + I_n)/2$.

\begin{center}

\begin{tabular}{||c||c|c||c|c||}
\hline
Model &
\multicolumn{2}{c||}{$\kappa_q = 0$}   &
\multicolumn{2}{c||}{$\kappa_q \neq 0$} \\ \cline{2-5}
    & $N - \Delta$ & $(p + n)/2$ & $N - \Delta$ & $(p + n)/2$ \\
\hline \hline
$A$ & $-107 \pm 9$  & $~-96$ & $-204 \pm 18$ & $-219$ \\
$B$ & $-114 \pm 9$  & $-120$ & $-193 \pm 16$ & $-219$ \\
$C$ & $-165 \pm 7$  & $-183$ & $-197 \pm ~9$ & $-219$ \\
$D$ & $-119 \pm 8$  & $-129$ & $-197 \pm 15$ & $-219$ \\ \hline
\end{tabular}

\end{center}

\newpage

\vskip 2cm

{\small \noindent
{\bf Table 3}. Values of the nucleon $DHG$ integrals $(I_p + I_n)/2$ and
$(I_p - I_n)/2$ (given in $\mu$barn), calculated for each model ($A -
D$) of the baryon wave functions using Eq. (\ref{eq:TOTAL}) with
$I_N^{bkg}$ given by Eq. (\ref{eq:INELASTIC}) and $I_N^{res}$ (Eq.
(\ref{eq:ELASTIC})) including only the $N - \Delta(1232)$ contribution.
The rows labelled $\langle \gamma_M \rangle \neq 1$ and $\langle
\gamma_M \rangle = 1$ correspond to the calculation of Eq.
(\ref{eq:INELASTIC}) with and without the relativistic factor
(\ref{eq:GAMMA}), respectively. The row labelled $DHG$ contains the
values obtained directly from the $DHG$ sum rule (\ref{eq:DHG}). The
results of the phenomenological analyses of Refs. \cite{KAR,WA} are
also reported.

\begin{center}

\begin{tabular}{||c||c|c||c|c||}
\hline
Model & \multicolumn{2}{c||}{$(I_p + I_n)/2$} &
        \multicolumn{2}{c||}{$(I_p - I_n)/2$} \\ \cline{2-5}
      & $\langle \gamma_M \rangle = 1$    & 
        $\langle \gamma_M \rangle \neq 1$ &
        $\langle \gamma_M \rangle = 1$    & 
        $\langle \gamma_M \rangle \neq 1$ \\ \hline \hline
 $A$  & $-223 \pm 18$ & $-213 \pm 18$ & $~16.5$ & $~~8.3$ \\ \hline
 $B$  & $-204 \pm 16$ & $-198 \pm 16$ & $~13.6$ & $~~6.9$ \\ \hline
 $C$  & $-200 \pm ~9$ & $-199 \pm ~9$ & $~~4.7$ & $~~2.9$ \\ \hline
 $D$  & $-205 \pm 15$ & $-203 \pm 15$ & $~~7.9$ & $~~5.9$ \\ \hline
\hline
 $DHG$               & \multicolumn{2}{c||}{$-219 $} &
                       \multicolumn{2}{c||}{$~14.7$} \\ \hline
 $Ref. ~ \cite{KAR}$ & \multicolumn{2}{c||}{$-222 $} &
                       \multicolumn{2}{c||}{$~-39 $} \\ \hline
 $Ref. ~ \cite{WA}$  & \multicolumn{2}{c||}{$-225 $} & 
                       \multicolumn{2}{c||}{$~-34 $} \\ \hline
\end{tabular}

\end{center}

\vskip 2cm

\begin{figure}[htb]

\epsfxsize=10cm \epsfig{file=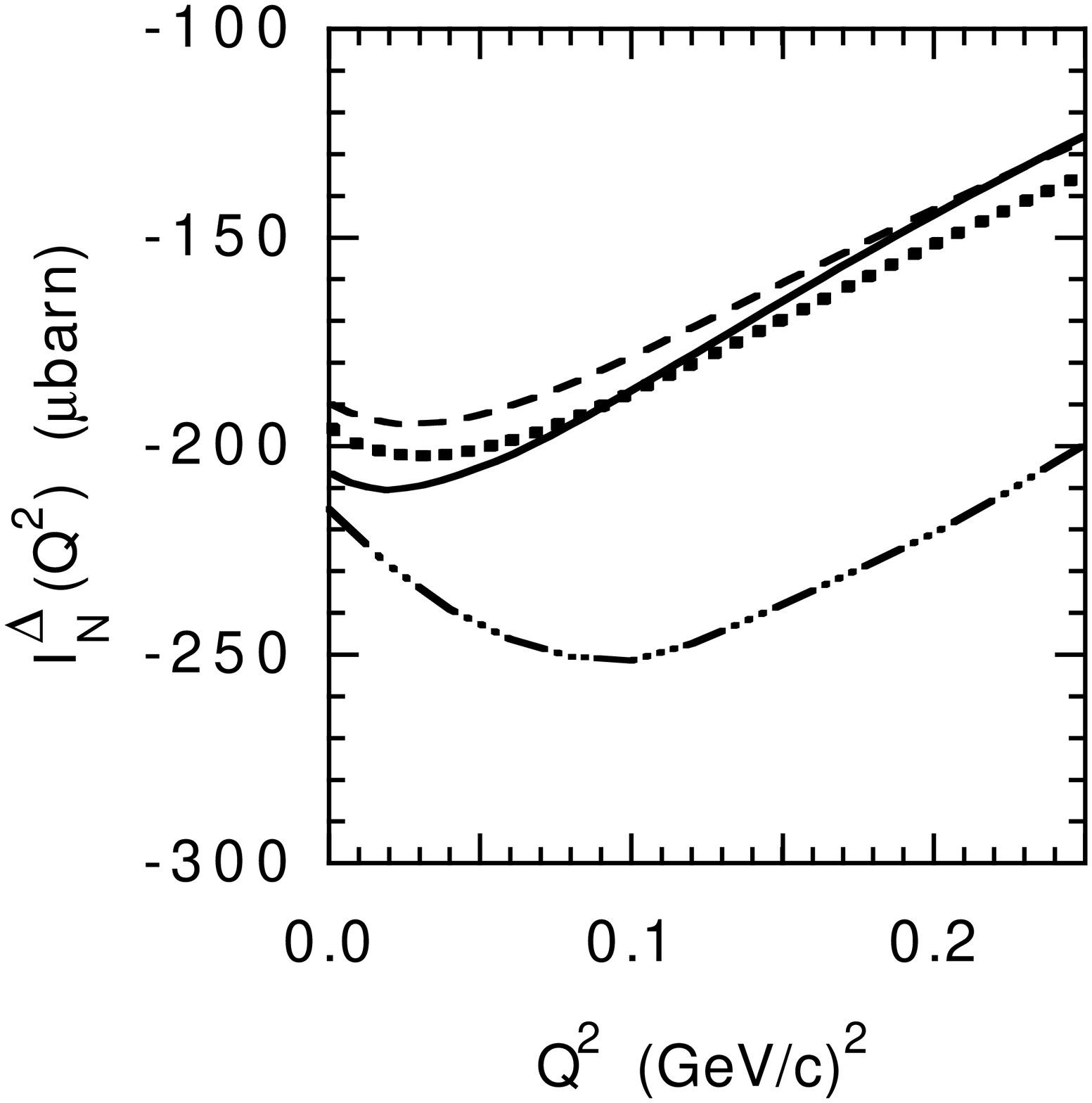}

\vskip -14cm

\parbox{10cm} \ $~~~~~~~~$ \ \parbox{6cm}{{\small \noindent
{\bf Figure 1}. The $N - \Delta(1232)$ contribution to the generalized
nucleon $DHG$ integral versus the squared four-momentum transfer $Q^2$.
The solid, dashed and dotted lines are our predictions calculated
within the models $A - C$, respectively, adopting the prescription
$III$ and using in Eq. (\ref{eq:CURRENT}) the $CQ$ form factors
determined by fitting pion and nucleon elastic data (see text). The
uncertainty due to the different prescriptions $I - III$ results to be
within $\sim 10 \%$. The dot-dashed line is the $N - \Delta(1232)$
contribution obtained within the relativized quark model of Ref.
\cite{LI}.}}

\end{figure}

\end{document}